\documentclass[journal=jacsat,manuscript=article]{achemso}
\usepackage{xcolor}
\usepackage{soul}
\usepackage[version=3]{mhchem} % Formula subscripts using \ce{}

\author{Liam S. Farrar\footnote{These authors contributed equally to this work.}}
\affiliation{Universit\'e Paris-Saclay, CNRS, Centre de Nanosciences et de Nanotechnologies (C2N), 91120 Palaiseau, France}
\author{Gaia Maffione$^{a}$}
\affiliation{Universit\'e Paris-Saclay, CNRS, Centre de Nanosciences et de Nanotechnologies (C2N), 91120 Palaiseau, France}
\author{Viet-Hung Nguyen$^{a}$}
\affiliation{Institute of Condensed Matter and Nanosciences, Université Catholique de Louvain (UCLouvain), 1348, Louvain-la-Neuve, Belgium}
\author{Kenji Watanabe}
\affiliation{Research Center for Electronic and Optical Materials, National Institute for Materials Science, 1-1 Namiki, Tsukuba 305-0044, Japan
}
\author{Takashi Taniguchi}
\affiliation{Research Center for Materials Nanoarchitectonics, National Institute for Materials Science,  1-1 Namiki, Tsukuba 305-0044, Japan}
\author{Jean-Christophe Charlier}
\affiliation{Institute of Condensed Matter and Nanosciences, Université Catholique de Louvain (UCLouvain), 1348, Louvain-la-Neuve, Belgium}
\author{Dominique Mailly}
\affiliation{Universit\'e Paris-Saclay, CNRS, Centre de Nanosciences et de Nanotechnologies (C2N), 91120 Palaiseau, France}
\author{Rebeca Ribeiro-Palau}
\affiliation{Universit\'e Paris-Saclay, CNRS, Centre de Nanosciences et de Nanotechnologies (C2N), 91120 Palaiseau, France}
\email{rebeca.ribeiro-palau@cnrs.fr}
\phone{+33 (0)1 70 27 06 92}

\title{Impact of the angular alignment on the crystal field and intrinsic doping of bilayer graphene/BN heterostructures}
\newcommand{\Dom}{\textcolor{black}}

\newcommand{\Reb}{\textcolor{black}}

\begin{document}


\begin{abstract}
  The ability to tune the energy gap in bilayer graphene makes it the perfect playground for the study of  the effects of internal electric fields, such as the crystalline field, which are developed \Reb{when other layered materials are deposited on top of it}. Here, we introduce a novel device architecture allowing a simultaneous control over the applied displacement field and the crystalline alignment between two materials. Our experimental  and numerical results confirm that the crystal field and electrostatic doping due to the interface reflect the 120$^{\circ}$ symmetry of the bilayer graphene/BN heterostructure and are highly affected by the commensurate state. These results provide an unique insight into the role of twist angle in  the development of internal crystal fields and intrinsic electrostatic doping in heterostructures. Our results highlight the importance of layer alignment, beyond the existence of a moir\'e superlattice, to understand the intrinsic properties of a heterostructure.
\end{abstract}


% \section{Introduction}
Van der Waals heterostructures have become the perfect playground for condensed matter physics, \Dom{especially} since the angular alignment between layers was added as a new knob to tune their properties. Using the right combination of materials and twist angle, very complex phenomena can be observed in these heterostructures, such as, superconductivity \cite{Cao2018Apr,Kim2022Jun,Yankowitz2019Mar,Guo2025Jan,Xia2024May}, Mott insulators \cite{Chen2019Mar}, anomalous quantum Hall effect\cite{Serlin2019Dec}, and more recently ferroelectricity \cite{zheng2020unconventional, niu2022giant, klein2023electrical, yan2023moire}. However, mastering all the parameters to  have an ultimate control of the properties of the heterostructure requires a deep understanding of the materials that compose the heterostructure and their interfaces. At these interfaces, crystal fields  effects result from the electrochemical  difference between layers, modifying the properties of the heterostructure  \cite{Rickhaus2019Oct}. Measuring the effect of the crystal field in heterostructure has only being addressed recently for fixed crystallographic alignments\cite{Rickhaus2019Oct}. This is mostly because investigating the angular variations will require the integration of dynamically rotatable van der Waals heterostructures \cite{ribeiro2018twistable, Inbar2023Feb} with techniques that allow to control, both, top and bottom gates, which implies an important technological challenge.

% \section{Results and discussion}

In this work, \Dom{we present a new device architecture that, in addition to rotational control of van der Waals heterostructures, provides control over both the top and bottom electrostatic gates.} We demonstrate the performance of this new architecture using Bernal-stacked bilayer graphene (BBG)  stacked between hexagonal boron nitride (BN). The control over both angular alignment and displacement field allows us to extract the angular dependence of the crystal electric field and intrinsic doping in a van der Waals heterostructure, pointing out the important role of layer alignment beyond the formation of moir\'e superlattices.

\begin{figure*}
    \centering
    \includegraphics[width=1\linewidth]{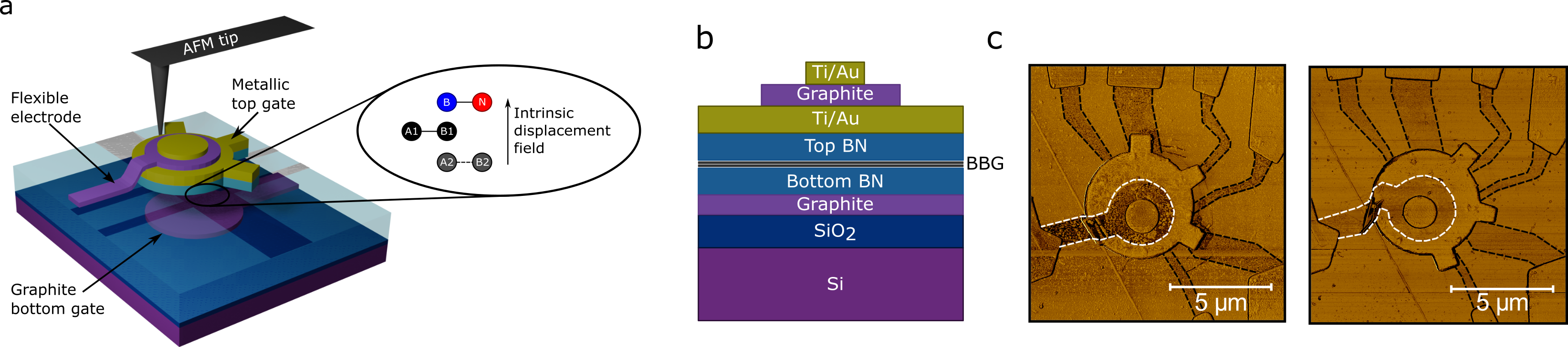}
    \caption{\textbf{Dual gated rotatable device architecture.} \textbf{a} and \textbf{b}, Illustration of the new device architecture (a) and its schematic cross-section (b). The zoom in (a) is meant to signal the BBG unit cell and the intrinsic displacement field. A1(2) and B1(2) represent carbon atoms in sublattice A and B of layer 1 (2), respectively. B (blue) and N (red) represent boron and nitrogen atoms, respectively. \textbf{c},  Atomic force microscopy images (phase signal) of a complete device before and after rotation, the graphite flexible electrode (white dashed line) and graphene Hall bar (black dashed lines) have been highlighted for clarity.}
    \label{fig:Figure_1}
\end{figure*}

Figures \ref{fig:Figure_1}a and b illustrate our novel device design, both as an overview (a) and its cross-section (b), Fig. \ref{fig:Figure_1}c shows atomic force microscope (AFM) images of a device in operation.
The device comprises two main components: firstly, a Hall-bar shaped BBG on a BN layer, which itself sits atop a pre-etched graphite electrode, separated from the BBG by a BN layer; secondly, a rotator made of BN and Ti/Au, connected via an etched graphite electrode.
The bottom graphite gate matches the rotator in diameter, and the graphite atop the rotator acts as a flexible van der Waals electrode for the application of a top gate voltage \cite{telford2018via}, creating a dual-gated region centered on the BBG Hall bar.
The Si substrate is highly doped in order to improve the contact resistances on the BBG far from the rotator by tuning the carrier density with a gate voltage \cite{ribeiro2019high}.
Full details of the fabrication procedures as well as optical images of the devices are provided in the Supplementary Note 1. It is important to mention that the bottom BN and graphene are intentionally misaligned via straight edge identification.

%It is important to mention that the bottom BN and graphene are intentionally misaligned via straight edge identification [reference probably]. Could also add:Although during the fabrication of the devices, the exact angular alignment is not known, it may be determined later on through techniques such as peak force microscopy [reference].

The rotation procedure is identical to our previous reports \cite{ribeiro2018twistable,arrighi2023non}, wherein an AFM tip is used to mechanically push the rotator while measuring simultaneously the resistance of the sample, thus providing {\it in situ} control of  the crystallographic alignment between the top BN and the BBG underneath. 
Overall, the device allows for rotational control within 140$^{\circ}$. In the AFM images of Fig. \ref{fig:Figure_1}c we can see examples where the BN handle is rotated by about 37$^{\circ}$ after being pushed with the AFM tip. In this report we present results obtained mainly in three samples: Sample 1 and Sample 2, which share the same bottom BN, and Sample 3 (see Supplementary Note 2).
The Hall bar widths $w$ and lengths $l$ are $w$ = $l$ = 1.5 $\mathrm{\mu m}$.
The carrier mobility of our samples ranged from 100,000 to 200,000 cm$^{2}$V$^{-1}$s$^{-1}$ for intermediate densities $\pm$0.65 x 10$^{12}$ cm$^{2}$ at $T<$ 40 K.
The mean free path was calculated to be $\approx$1.5 $\mathrm{\mu m}$ for the same carrier density below $T<$ 40 K, reflecting the ballistic character of the charge transport for all of our devices. 

\begin{figure*}
    \centering
    \includegraphics[width=1\linewidth]{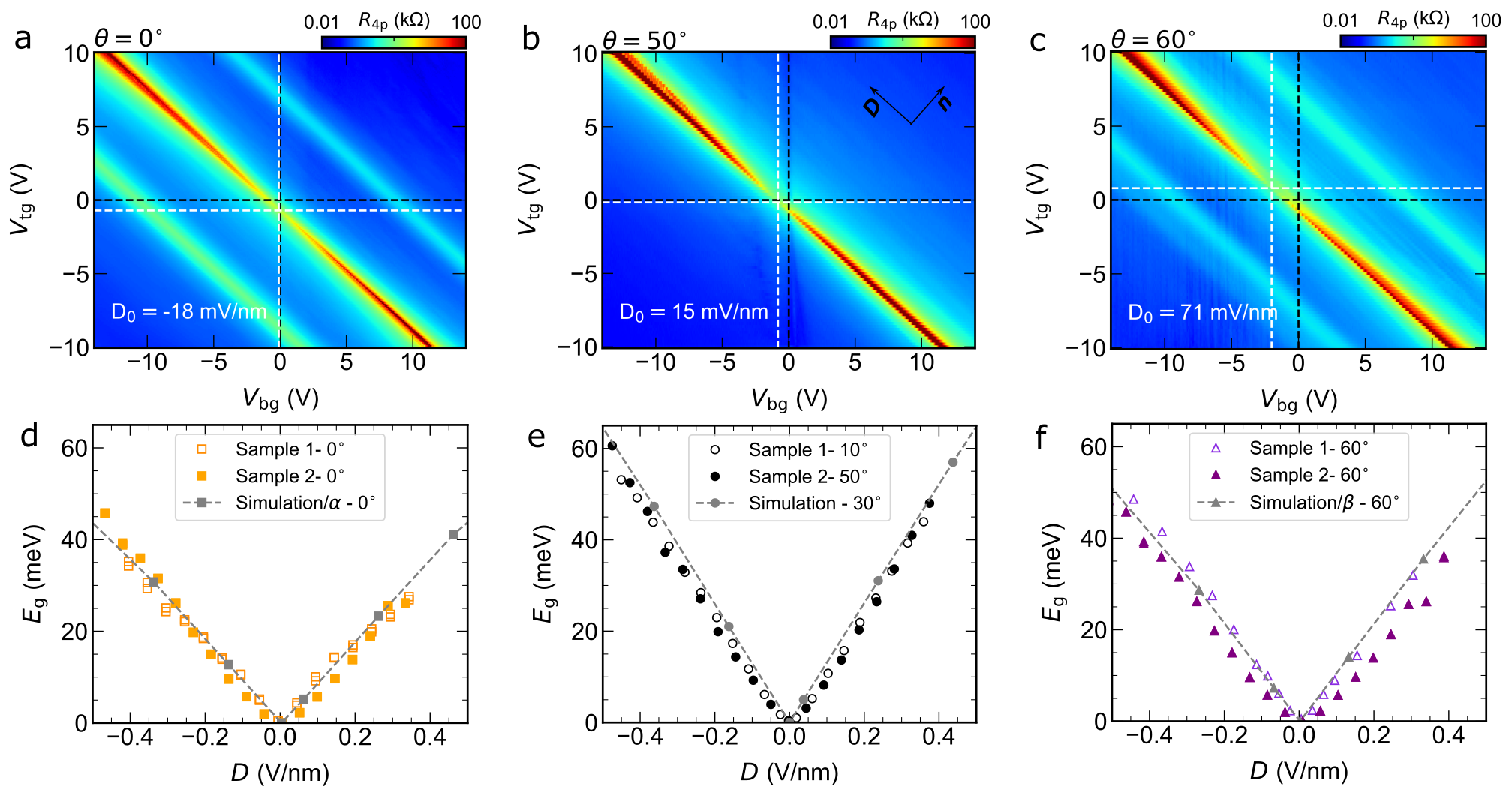}
    \caption{\textbf{Dual gate effect at different rotational angles.} Four-terminal resistance, $R_{\mathrm{4P}}$, as a function of top ($V_{\mathrm{tg}}$) and bottom ($V_{\mathrm{bg}}$) gate voltages at $\theta_{\mathrm{top}}=0^{\circ}$ \textbf{a},   $\theta_{\mathrm{top}}=50^{\circ}$ \textbf{b}, and $\theta_{\mathrm{top}}=60^{\circ}$ \textbf{c} of alignment between the top BN and the BBG for Sample 2 taken at $T=10$ K. Black dashed lines represent the $V_{\mathrm{tg}}= 0$ V and $V_{\mathrm{bg}}=0$ V lines. White dashed lines show the position of the lowest value on resistance of the CNP. \textbf{d}-\textbf{f}, Energy gap, $E_{\mathrm{g}}$, as a function of the total displacement field, $D$, for three alignments: $\theta_{\mathrm{top}}=0^{\circ}$ \textbf{d},  $\theta_{\mathrm{top}}=10^{\circ}/50^{\circ}$ \textbf{e}, and $\theta_{\mathrm{top}}=60^{\circ}$ \textbf{f},  for Sample 1 and Sample 2. The data points are extracted from the Arrhenius plot in the thermally activated regime. Gray points with dashed lines are the result of numerical simulations divided by a factor $\alpha=1.4$ and $\beta=1.2$ for 0$^{\circ}$ and 60$^{\circ}$, respectively. See main text. }
    \label{fig:Figure_2}
\end{figure*}

To demonstrate the efficiency of our dual gated rotatable devices we show in Figs. \ref{fig:Figure_2}a-c measurements of the four-probe resistance of sample 2 at 10 K as a function of $V_{\mathrm{bg}}$ and $V_{\mathrm{tg}}$ at three different angles: $\theta_{\mathrm{top}}=0^{\circ}$ (a), $\theta_{\mathrm{top}}\approx50^{\circ}$ (b), where the moir\'e has little to no influence, and $\theta_{\mathrm{top}}=60^{\circ}$ (c).

%%%%%%%%%%%%%% HERE

In BBG, when the sign of $V_{\mathrm{bg}}$ and  $V_{\mathrm{tg}}$ are opposite we observe a rapidly increasing $R_{\mathrm{4P}}$ at the charge neutrality point (CNP). This maximum resistance for the three alignments, traces a straight line, with a slope determined by the ratio of the capacitive coupling of each layer $C_{\mathrm{tg}}$/$C_{\mathrm{bg}}$ which is in turn directly related to the thickness ratio between the top and bottom BN layers \cite{taychatanapat2010electronic}.
Along this diagonal line, the resistance strongly increases, a clear sign that a band energy gap is opening, as the magnitude of the displacement field $|D|$  increases \cite{Zhang2009Jun, taychatanapat2010electronic,icking2022transport}.  The total displacement field,  $D$,  and total carrier density,  $n_{\mathrm{T}}$, of the system can  be expressed as:

\begin{equation}
    D = D_{\mathrm{A}}-D_{\mathrm{0}}=\frac{e}{2\epsilon_{0}}[C_{\mathrm{tg}}V_{\mathrm{tg}}- C_{\mathrm{bg}}V_{\mathrm{bg}}] -D_{\mathrm{0}}
\end{equation} 

\noindent and 
\begin{equation}
  n_{\mathrm{T}} = n - n_{\mathrm{0}}= [C_{\mathrm{tg}}V_{\mathrm{tg}} + C_{\mathrm{bg}}V_{\mathrm{bg}}] - n_{\mathrm{0}},
\end{equation}

\noindent where $D_{\mathrm{A}}$ is the applied displacement field, $D_{\mathrm{0}}$ is the residual or intrinsic displacement field of the sample, $n_{\mathrm{0}}$ is the residual carrier density or intrinsic doping of the bilayer, $C_{\mathrm{b(t)g}}=\epsilon_{0}\epsilon/d_{\mathrm{b(t)}} e$ is the bottom (top) gate capacitive coupling, directly extracted from the Hall effect, $\epsilon$ is the BN dielectric constant, $d_{\mathrm{b(t)}}$ is the BN thickness of the bottom (top) gate, and $\epsilon_{0}$ is the vacuum permittivity. 

The behavior of the two aligned positions $\theta_{\mathrm{top}}=0^{\circ}$ and $\theta_{\mathrm{top}}=60^{\circ}$, Figs. \ref{fig:Figure_2}a and c, are dissimilar due to the 120$^{\circ}$ symmetry of the BN/BBG heterostructucture \cite{arrighi2023non} and are distinguished by the relative heights of the room temperature four-probe resistance measurements (see Supplementary Note 3 for further details). At low temperatures, their alignment is indicated by the presence of satellite peaks which appear symmetrically in carrier density around the CNP and occur due to the emergence of satellite Dirac points induced by scattering from the moir\'e superlattice potential \cite{yankowitz2012emergence}.  The moir\'e wavelength $\lambda$ is accurately determined by magneto-transport measurements and corresponds to $\lambda$ = 14.2 $\pm$ 0.1 nm and $\lambda$ = 14.0 $\pm$ 0.1 nm for the 0$^{\circ}$ and 60$^{\circ}$ alignments, respectively (see Supplementary Note 4 for further details).

Notice that, for the aligned positions, contrary to the CNP, the resistance of the satellite peaks  do not significantly change with displacement field, Figs. \ref{fig:Figure_2}a and c. This means that these peaks are not affected by the breaking of inversion symmetry created by the displacement field.

To determine the residual or intrinsic displacement field, $\mathrm{D_{\mathrm{0}}}$, we follow the resistance at the CNP as a function of both electrostatic gates. As explained before, the CNP can be clearly seeing as the red diagonal line in Figs. \ref{fig:Figure_2}a-c, which indicate a strong resistance. The lowest resistance value along this diagonal line is the point where the applied $V_{\mathrm{bg}}$ and $V_{\mathrm{tg}}$ fully close the energy band gap. We have marked it by the crossing of white dashed lines in Figs. \ref{fig:Figure_2}a-c. At this point the applied displacement field, $D_{\mathrm{A}}$, compensates the intrinsic displacement field, $D_{\mathrm{0}}$, of the sample. The origin of the intrinsic displacement field, as well as the residual doping, in single gated devices has been attributed to contaminants and residues on the surface of the device, however, it is more difficult to spot its origin in the case of double gated devices as ours, see discussion below.

The energy gap, $E_{\mathrm{g}}$, induced by the applied displacement field can be determined via the temperature dependence of the CNP resistance. In the thermally activated regime, the resistance at the CNP decays exponentially as  $R_{\mathrm{CNP}} \propto \exp(E_{\mathrm{g}}/(2k_{\mathrm{B}T}))$, where  $k_{\mathrm{B}}$ is the Boltzmann's constant and $T$ the temperature (Supplementary Note 5).  In Figs. \ref{fig:Figure_2}d-f we plot the dependence of the energy gap with the total displacement field, $D$. All the curves follow an approximately linear dependence of the energy gap, as previously reported  \cite{icking2022transport,taychatanapat2010electronic,Iwasaki2022Oct}. However, they do not have the same slope depending on the angular alignment. For the misaligned cases, Fig. \ref{fig:Figure_2}e, the linear behavior is highly symmetric with respect to positive and negative displacement field, in agreement with previous reports \cite{icking2022transport}.  On the other hand, for the aligned position, Figs. \ref{fig:Figure_2}d and f, the energy gap is slightly smaller.  It is important to notice that, contrary to previous reports \cite{Iwasaki2022Oct}, and consistent with our earlier results \cite{arrighi2023non}, we do not observe any thermal activated behavior for the satellite peaks at any value of the displacement field.

\begin{figure*}
    \centering
    \includegraphics[width=1\linewidth]{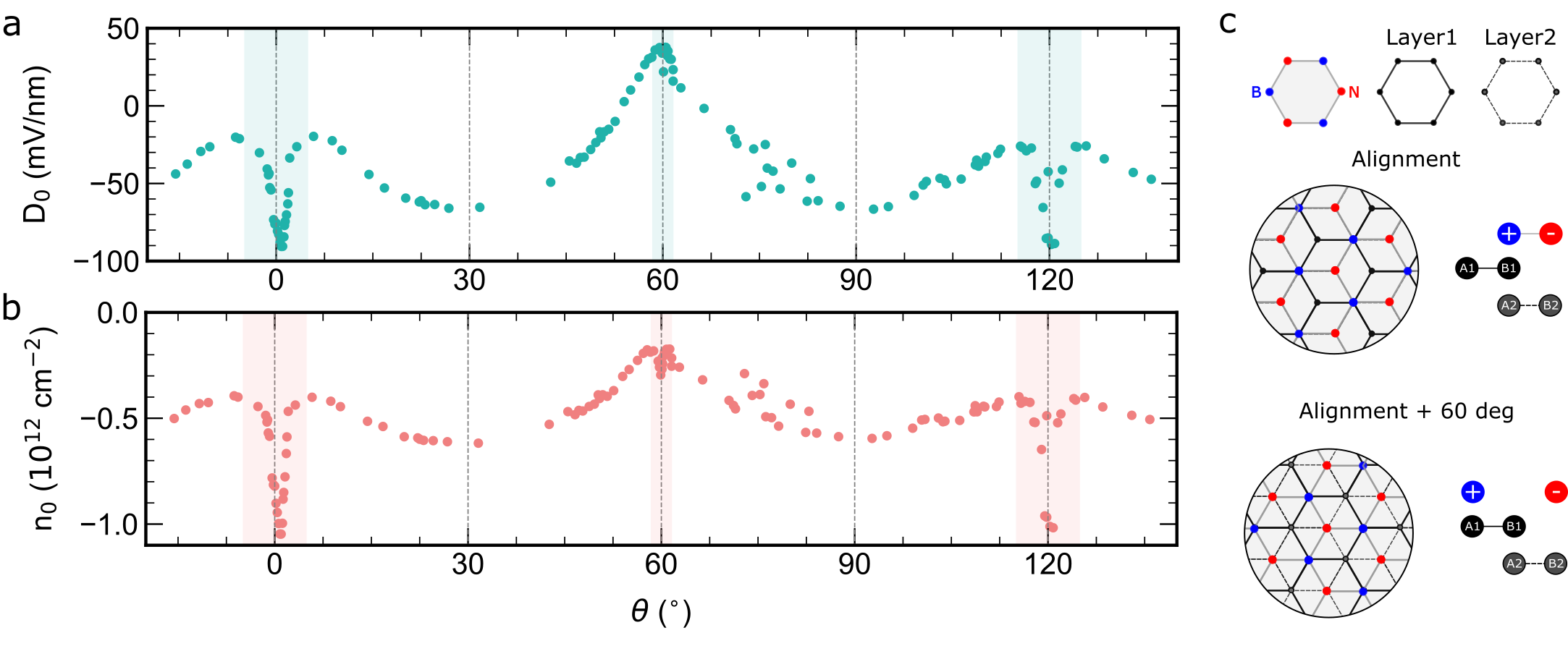}
    \caption{\textbf{Intrinsic doping and crystal field as a function of angle}.   \textbf{a}, Angular dependence of the internal displacement field generated by the presence of the top and bottom BN. \textbf{b}, Angular dependence of the intrinsic doping of the system.Shadow areas around 0$^{\circ}$, 60$^{\circ}$ and 120$^{\circ}$ represent the $\pm5^{\circ}$ ($\pm2^{\circ}$) segment where a deviation from the expected behavior is observed for 0$^{\circ}$ (60$^{\circ}$). Measurements taken for sample 3. \textbf{c}, Schematic representation of the two crystallographic alignments separated by 60$^{\circ}$. Here positive charge accumulation on B (blue) atoms and negative charge accumulation on N (red) atoms is represented by ($+$) and ($-$) signs respectively.}
    \label{fig:Figure_3}
\end{figure*}

Let's discuss these features. The energy band gap, $E_{\mathrm{g}}$ is given by \cite{min2007ab, mccann2006asymmetry}

\begin{equation}
    E_{\mathrm{g}} = \frac{|\Delta|}{\sqrt{1 + (\Delta/\gamma_{1})^2}},
    \label{Eqn: Eqn3}
\end{equation}

\noindent where $\gamma_{1}$ is the interlayer coupling strength between the graphene layers and $\Delta$ is the onsite potential difference expressed as \cite{mccann2006asymmetry, jung2014accurate}:

\begin{equation}
    \Delta = \frac{d_0 e \widetilde{D}}{\varepsilon_0 \varepsilon_{\mathrm{z}}}+\frac{d_0 e^2}{2 \varepsilon_0 \bar{\varepsilon}} \delta n(\Delta).
    \label{Eqn: Eqn4}
\end{equation}

The first term of Eq. \ref{Eqn: Eqn4} results from  the external (applied or intrinsic) displacement field, where $\widetilde{D}/\epsilon_{0}=D$ is the total displacement field, $d_{0}$ is the interlayer spacing between the graphene layers, $e$ is the elementary charge, and $\varepsilon_{\mathrm{z}}$, $\bar{\varepsilon}$ are effective dielectric constants which vary depending on the model used \cite{Slizovskiy2021Jul,McCann2013Apr}. In our numerical simulation we use $\varepsilon_{\mathrm{z}}=2.5$ \cite{Slizovskiy2021Jul}.
The second term of Eq. \ref{Eqn: Eqn4} results from the difference between the charge carrier densities in the upper $n_{1}$ and lower $n_{2}$ layers $\delta n = n_{2} - n_{1}$.
$\delta n$ itself depends also on the value of $\Delta$ and as such requires a self-consistent solution which, for the misaligned case, is presented elsewhere \cite{Slizovskiy2021Jul,McCann2013Apr,icking2022transport}. 

Intuitively, the difference observed for the energy gap between the aligned and misaligned cases could be explained as an effect of the commensurate state near crystallographic alignment, \Reb{a local enlargement of the graphene lattice constant due to its van der Waals interaction with BN} \cite{woods2014commensurate,arrighi2023non}. However, we have performed numerical simulations (same numerical methods as in \cite{arrighi2023non}) and the effects of the commensurate state in the energy gap seems only to be relevant at larger displacement field (see Supplementary Note 6). 

By directly comparing the result of our numerical simulations with the experimental results of the energy gap as a function of displacement field, Figs. \ref{fig:Figure_2}d-f,  we remark that the data from numerical simulations, for both aligned positions 0$^{\circ}$ and 60$^{\circ}$, \Reb{differ} by a factor of $\alpha=1.4$ and $\beta=1.2$. The origin of this discrepancy might come from the second term of equation eq. \ref{Eqn: Eqn4} since the $\delta n(\Delta)$ parameter is not included in our numerical simulation given the complexity of its implementation in a large moir\'e cell. This term seems to be negligible at the misaligned position since our numerical simulations fit very well the experimental data.

\Reb{We can extract experimentally the angular dependence of $\delta n_{0}$, the difference between the charge carrier densities in the upper and lower layers without any applied displacement field, Fig. \ref{fig:Figure_3}a, by calculating the intrinsic displacement field. For this, we performed resistance measurements around the CNP, by varing $V_{\mathrm{bg}}$, for fixed values of $V_{\mathrm{tg}}$ to obtain the values of $V_{\mathrm{tg}}$ and $V_{\mathrm{bg}}$ for which the height of the CNP is the lowest, reflecting the  closing of the energy gap (for details see Supplementary Note 7). These experiments are performed at room temperature to achieve a high angular resolution, with several points confirmed at low temperature to ensure the reported phenomena remain stable across temperatures (for details, see Supplementary Note 7). Fig. \ref{fig:Figure_3}a shows the angular dependence of the intrinsic displacement field, $D_{\mathrm{0}}$, which can be understood as the crystal field generated by the atomic configuration of the upper and bottom BN layers on the graphene layer.  In these measurements we observe a strong change at 0$^{\circ}$ occurring in a range of $\pm5^{\circ}$ and a change in the opposite direction for 60$^{\circ}$ of alignment that saturates for a range of angle of $\pm2^{\circ}$. This indicates that the alignment of the BN layers to the graphene also plays a major role since in the commensurate state strong local strains in graphene are induced by its interactions with BN affecting therefore the intrinsic displacement field.}

\Reb{The measurements described above also allow us to extract the angular dependence of the intrinsic doping, $n_{0}$. In Fig. \ref{fig:Figure_3}b we can see the intrinsic density, for misaligned angles, varies slowly with the angular alignment until it reaches the aligned positions. At these aligned positions, 0$^{\circ}$ and 60$^{\circ}$, the behavior is opposite, increasing  for 0$^{\circ}$ and decreasing for 60$^{\circ}$, this change is much more pronounced in the case of 0$^{\circ}$. The physical origin of this particular dependence of the intrinsic doping with the angular alignment is difficult to elucidate. However, since it is much more pronounced at the aligned positions its origin can be intuitively assigned to the strain generated by the commensurate state\cite{arrighi2023non}.}

\begin{figure}
    \centering
    \includegraphics[width=1\linewidth]{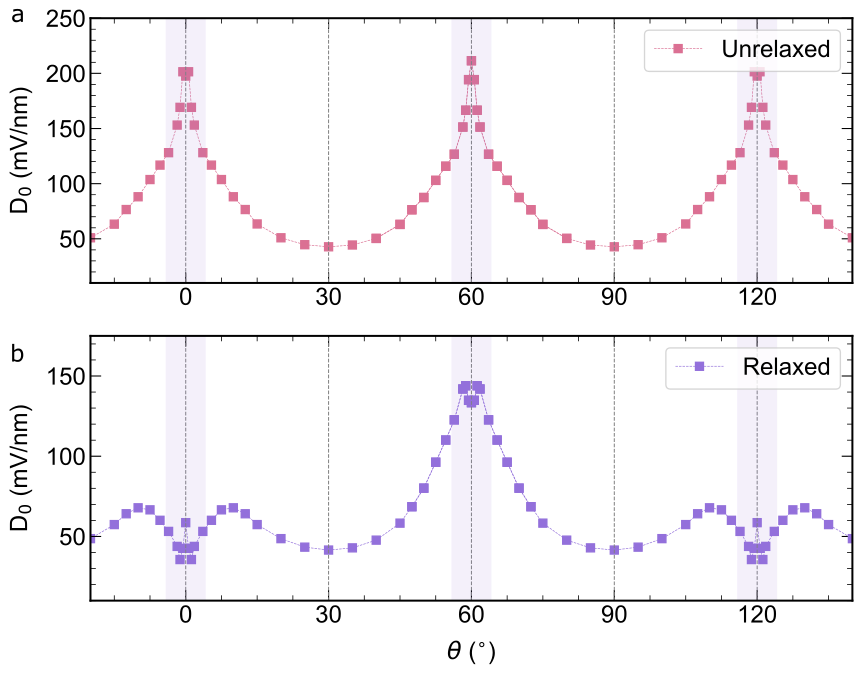}
    \caption{\textbf{Crystal field numerical simulation.}  Intrinsic displacement field as a function of the angular alignment between bilayer graphene and a BN layer for a system without atomic relaxation \textbf{a} and one with atomic relaxation \textbf{b}.}
    \label{fig:Figure_4}
\end{figure}

In order to understand this non-trivial dependence, we can start by comparing our experimental results with numerical simulations. As in the experiment, for the numerical simulations we have extracted the displacement field  needed in order to close the energy gap of the electronic band structure for each angular alignment, from now we will refer to this as the crystal field, since it is a natural electric field generated by the atomic configuration of our system. 

\Reb{To determine the mentioned crystal field, we first calculate the energy gap of the system for a given angular alignment without applied displacement field and then we numerically apply a displacement field in an opposite direction, {\it i.e.}, in order to cancel the potential difference of graphene layers and therefore closing the energy gap. In our simulations, the applied displacement field is simply computed by adjusting the onsite energies of each graphene layer, i.e., $\epsilon_{\mathrm{Carbon}} = \pm\Delta/2$, in the Hamiltonian of ref. \cite{arrighi2023non}.}

We can start by discussing the unrelaxed case, Fig.\ref{fig:Figure_4}a, for which we can see that the crystal field is minimum at 30$^{\circ}$ and 90$^{\circ}$ and increases strongly close to the aligned positions {\it i.e.}, 0$^{\circ}$, 60$^{\circ}$ and 120 $^{\circ}$. Note that the minimum value of the displacement field is not zero because we have simulated a system with only one BN layer on top of the bilayer graphene. The strong increase of the crystal field at the aligned position is due to the combination of two factors: {\it i)} the electric field generated by the boron and nitrogen atoms by their proximity to the atoms A1 and B2 of the BBG (crystal field effect), see Fig. \ref{fig:Figure_3}c, and {\it ii)} the moir\'e superlattice, which modifies the electronic band structure, even in the unrelaxed case. By using the numerical methods explained in \cite{arrighi2023non} we relax the atomic structure, Fig. \ref{fig:Figure_4}b. This atomic relaxation will give rise to the stretching of the graphene lattice inside the moir\'e cell and the formation of wrinkles of unstretched graphene around these cells.

For the relaxed numerical simulations near crystallographic alignment the crystal field is highly impacted by the moir\'e superlattice, as predicted in \cite{Tepliakov2021May}. However, we can see that the impact for 0$^{\circ}$ and 60$^{\circ}$ has opposite behavior, which can be explained by looking at the unit cell of BBG for both aligned positions 0$^{\circ}$ and 60$^{\circ}$, Fig. \ref{fig:Figure_3}c. In the commensurate state, at the inner part of the moir\'e cell the BBG atoms are arranged in a BA stacking, between top layer of graphene and the BN layer. Here boron atoms are preferentially sitting on the carbon atoms, since this is the most energetically favorable configuration \cite{Jung2015Feb}. We have demonstrated in our previous report \cite{arrighi2023non} that the difference in the stacking for both alignment give rise to differences in the in-plane atomic displacement and therefore to the electronic transport characteristics.

In the case of the crystal field, the difference between boron ($+$) and nitrogen ($-$) will generate a different charge accumulation between the sites A1 and B2, inducing a different crystal field. This crystal field is expected to be stronger in one of the aligned cases because of the configuration of nitrogen and boron atoms, which also creates a larger $\delta n_{0}$. As this is observed within the same sample, we can disregard any effect of  spurious doping, providing further evidence of the non-identical nature of the two moir\'e in BBG/BN. It is important to note that our numerical simulations do not consider the bottom BN which induces an offset in Fig. \ref{fig:Figure_3}a, explaining the asymmetry with respect to zero crystal field. This can be seen in the Supplementary Note 7 where the general behavior of the curve for a different sample is the same the curve is offset.

% \subsection{Conclusions}

To conclude, we have developed a new device architecture that enables the  rotation control of a van der Waals heterostructure while allowing for a simultaneous control of both  bottom and  top gates.
We have used this device to reveal further details about effects of the moir\'e superlattice such as its impact on the evolution of the band gap by an external displacement field. We have measured for the first time the evolution of the atomic crystal field with angular alignment  in a van der Waals heterostructure and revealed the strong effects of the commensurate state.  The commensurate state in aligned bilayer graphene/BN structures is, as predicted by our numerical simulations, 120$^{\circ}$ periodic. These results highlight the importance of minding the layer alignment of each layer, beyond the existence of a moir\'e superlattice, to understand and simulate the intrinsic properties of a van der Waal heterostructure.

\begin{acknowledgement}

The authors acknowledge discussions with Herv\'e Aubin, Ulf Gennser, Nicolas Leconte and Cory Dean.
 This work was done within the C2N micro nanotechnologies platforms and partly supported by the RENATECH network and the General Council of Essonne. This work was supported by: ERC starting grant N$^{\circ}$ 853282 - TWISTRONICS (R.R-P.), the DIM-SIRTEQ project TOPO2D, the DIM QuanTIP project Q-MAG and IQUPS. R.R.-P. and J.-C.C. also acknowledge the Flag-Era JTC project TATTOOS (N$^{\circ}$ R.8010.19) and the Pathfinder project ``FLATS'' N$^{\circ}$ 101099139. V.H.N and J.-C.C. also acknowledge fundings from the F\'ed\'eration Wallonie-Bruxelles through the ARC project ‘‘DREAMS’’ (N° 21/26-116), from the EOS project ‘‘CONNECT’’ (N° 40007563), and from the Belgium F.R.S.-FNRS through the research project (N$^{\circ}$ T.029.22F). Computational resources were provided by the CISM supercomputing facilities of UCLouvain and the CÉCI consortium funded by F.R.S.-FNRS of Belgium (N$^{\circ}$ 2.5020.11). K.W. and T.T. acknowledge support from the JSPS KAKENHI (Grant Numbers 21H05233 and 23H02052) and World Premier International Research Center Initiative (WPI), MEXT, Japan.

\end{acknowledgement}

\begin{suppinfo}

Additional experimental details, materials, methods, and theoretical model (pdf)

\end{suppinfo}

\bibliography{achemso-demo}

\section*{TOC Graphic}

\begin{figure}
    \centering
    \includegraphics[width=1\linewidth]{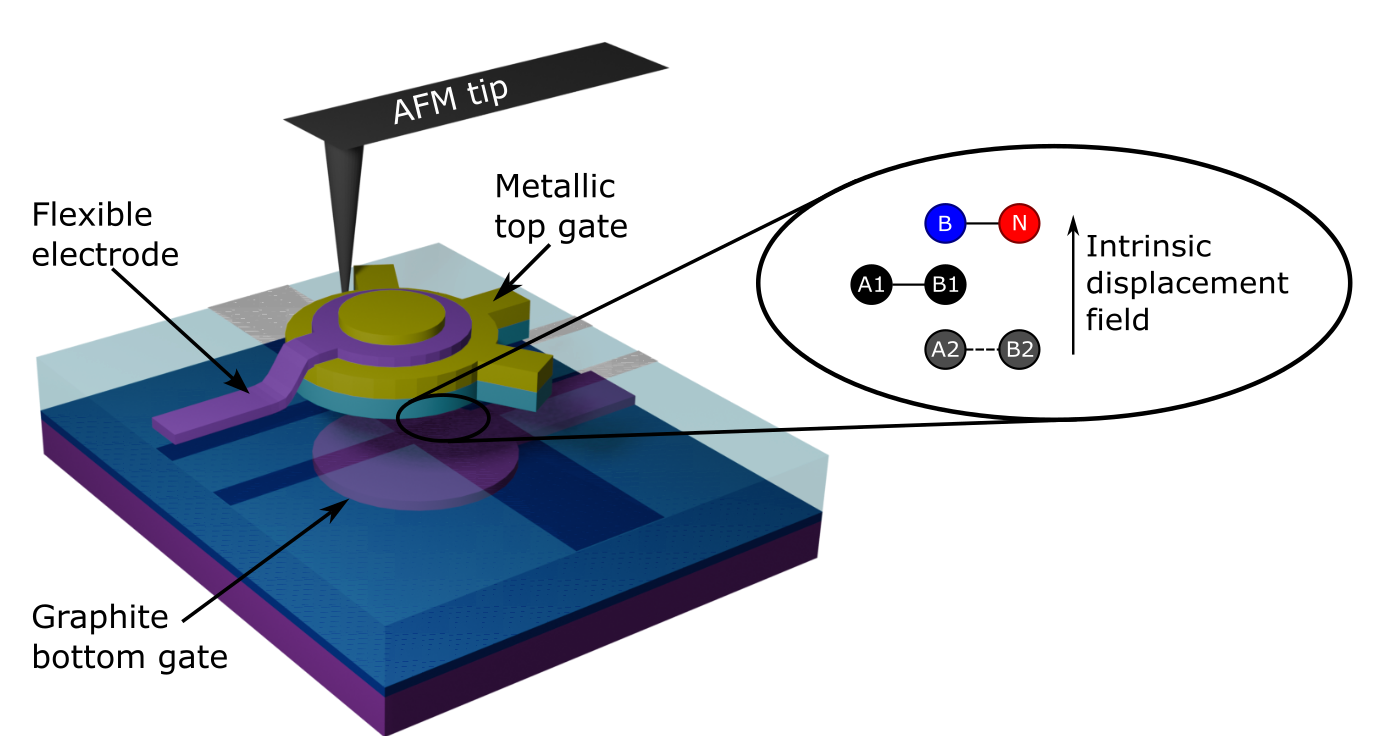}
    
    \label{fig:TOC}
\end{figure}

\end{document}